\documentclass[english]{article}
\usepackage[T1]{fontenc}
\usepackage{verbatim}
\usepackage{float}
\usepackage{amsthm}
\usepackage{amsmath}
\usepackage{amssymb}
\usepackage{graphicx}
\usepackage{color}
\usepackage{url}
\usepackage{authblk}
\usepackage{xr}

\usepackage[colorlinks=true,bookmarks=false,linkcolor=blue,urlcolor=blue,citecolor=blue,breaklinks=true]{hyperref} 

\makeatletter

\floatstyle{ruled}
\newfloat{algorithm}{tbp}{loa}
\providecommand{\algorithmname}{Algorithm}
\floatname{algorithm}{\protect\algorithmname}

\numberwithin{equation}{section}
\numberwithin{figure}{section}
\theoremstyle{plain}

\theoremstyle{definition}

\theoremstyle{remark}

\theoremstyle{plain}

\newtheorem*{lem*}{Lemma}

\theoremstyle{remark}

\theoremstyle{plain}

\theoremstyle{plain}

\@ifundefined{showcaptionsetup}{}{%
 \PassOptionsToPackage{caption=false}{subfig}}
\usepackage{subfig}
\makeatother

\usepackage{babel}
\providecommand{\claimname}{Claim}
\providecommand{\definitionname}{Definition}
\providecommand{\lemmaname}{Lemma}
\providecommand{\remarkname}{Remark}
\providecommand{\theoremname}{Theorem}
\providecommand{\corollaryname}{Corollary}
\providecommand{\propositionname}{Proposition}

\newcommand{\reals}{\mathbb{R}}
\newcommand{\RL}{\mathbb{R}^L}

\newcommand{\calO}{O}
\newcommand{\dist}{\operatorname{dist}}
\newcommand{\Var}{\operatorname{Var}}
\newcommand{\EE}{\mathbb{E}}

\usepackage{spconf}

\title{Heterogeneous multireference alignment: a single pass approach}

\name{Nicolas Boumal,$^{1,2}$ \quad  Tamir Bendory,$^{2}$ \quad Roy R.\ Lederman,$^{2}$ \quad Amit Singer $^{1,2}$ \thanks{The authors were partially supported by Award Number R01GM090200 from the NIGMS, FA9550-17-1-0291 from AFOSR, Simons Investigator Award and Simons Collaboration on Algorithms and Geometry from Simons Foundation, and the Moore Foundation Data-Driven Discovery Investigator Award. NB is partially supported by NSF grant DMS-1719558.}}

\address{Mathematics Department$^1$ and PACM$,^2$ Princeton University, Princeton, NJ, USA}

\begin{document}
\ninept
\maketitle
\begin{abstract}
Multireference alignment (MRA) is the problem of estimating a signal from many noisy and cyclically shifted copies of itself. In this paper, we consider an extension called \emph{heterogeneous} MRA, where $K$ signals must be estimated, and each observation comes from one of those signals, unknown to us. This is a simplified model for the heterogeneity problem notably arising in cryo-electron microscopy. We propose an algorithm which estimates the $K$ signals without estimating either the shifts or the classes of the observations. It requires only one pass over the data and is based on low-order moments that are invariant under cyclic shifts. Given sufficiently many measurements, one can estimate these invariant features averaged over the $K$ signals. We then design a smooth, non-convex optimization problem to compute a set of signals which are consistent with the estimated averaged features. We find that, in many cases, the proposed approach estimates the set of signals accurately despite non-convexity, and conjecture the number of signals $K$ that can be resolved as a function of the signal length $L$ is on the order of $\sqrt{L}$.
\end{abstract}

\begin{keywords}
Multireference alignment, bispectrum, non-convex optimization, expectation-maximization, Gaussian mixture models, cryo-EM, heterogeneity
\end{keywords}


\section{Introduction}

Multireference alignment (MRA) seeks to estimate a signal from numerous noisy and cyclically shifted copies of itself. This problem serves as a model for scientific problems in structural biology~\cite{diamond1992multiple,theobald2012optimal,park2011stochastic,park2014assembly}, radar~\cite{zwart2003fast,gil2005using} and image processing~\cite{dryden1998statistical,foroosh2002extension,robinson2009optimal}. Algorithmic and statistical properties of MRA have been analyzed in~\cite{bendory2017bispectrum,bandeira2017optimal,perry2017sample,bandeira2014multireference,abbe2017sample}.

In this paper, we consider \emph{heterogeneous} MRA, where more than one signal must be estimated from noisy and shifted observations. This investigation is motivated in part by applications in single particle cryo-electron microscopy (cryo-EM) and X-ray free electron lasers (XFEL). These imaging techniques are used to map the three-dimensional structure of molecules at near-atomic resolution from either two-dimensional noisy tomographic  projections (cryo-EM) or diffraction images (XFEL),
taken at unknown viewing directions~\cite{bartesaghi20152,mcneil2010x,frank2006three}. 
It is typical in those applications to acquire a large number of very noisy observations. 
Heterogeneity arises when more than one type of molecule (or conformation) appears in a sample.
{Heterogeneous} MRA is a simplified one-dimensional model of this situation, where the rotations are replaced by cyclic shifts. The tomographic projection from three-dimensional objects to two-dimensional images is not modeled in this formulation of MRA.

One of the opportunities in cryo-EM and XFEL---com\-pared to X-ray crystallography---is that they potentially allow to estimate multiple conformations of molecules observed together in heterogeneous mixtures. Achieving this capability is one of the key technological challenges~\cite{nogales2016development}, and the subject of many recent works~\cite{schwander2014conformations,frank2016continuous,katsevich2015covariance,anden2015covariance,lederman2016representation,aizenbud2016max,lederman2017continuously}.

This motivates us to consider heterogeneity in MRA, where $K \geq 1$ signals must be recovered from unlabeled, shifted, noisy observations. It has been shown in~\cite{perry2017sample} that such a mixture of signals can be estimated from fifth-order moments using a tensor decomposition algorithm. In this paper, we propose a method which, empirically, estimates all $K$ signals simultaneously using only the third-order moments, same as what is necessary for the homogeneous case~\cite{bandeira2017optimal}. 

In a nutshell, following~\cite{bendory2017bispectrum}, for each observation, we compute features (moments) which are invariant under cyclic shift. Namely, we compute the mean, power spectrum and bispectrum. Averaging these features over all observations yields an estimator for the averaged invariant features. We then set up a smooth, non-convex optimization problem designed to recover the unknown signals from the averaged features.
Our numerical study demonstrates that this approach performs well for a broad range of parameters, with random initialization, despite the non-convexity of the optimization problem.

Importantly, our approach bypasses the need to estimate the latent variables of the problem, namely, the unknown shifts and classes of the individual observations. Furthermore, it naturally works in single-pass streaming mode and is parallelizable. 
We show that for large data and low signal-to-noise ratio (SNR), this can be about as accurate and much faster than a standard alternative, namely, expectation maximization~(EM).

We note that signal estimation based on invariant features is an old idea in signal processing~\cite{sadler1992shift} and cryo-EM \cite{kam1980reconstruction,marabini1996new}. Furthermore, the idea of estimating more than one object from mixed measurements was recently used for Gaussian mixtures~\cite{hsu2013learning,anandkumar2014tensor} and for mixed low-rank matrix sensing~\cite{strohmer2017painless}.
Our main contribution is to demonstrate how a non-convex optimization approach to heterogeneous MRA resolves multiple signals in a single pass over the data at low SNR.


\section{MRA without heterogeneity} \label{sec:mra}

We begin by introducing the homogeneous MRA model.
Let $x \in \RL$ be the unknown signal and let $R_{r}$ be the cyclic shift operator: $(R_{r}x)[n] = x[n-r]$, with all indices considered modulo $L$.
We are given $N$ measurements:
\begin{align}
	y_j = R_{r_j}x + \varepsilon_j, \quad j = 1, \dots, N,
	\label{eq:mra}
\end{align}
where $\varepsilon_j \sim \mathcal{N}(0, \sigma^2 I)$ is i.i.d.\ white Gaussian noise. Our goal is to estimate $x$ up to shift in a high-noise regime where the unknown shifts $r_j$ could not be recovered reliably even if $x$ were known (see Cram\'er--Rao lower bounds in~\cite{aguerrebere2016fundamental}).

Following recent work~\cite{bendory2017bispectrum}, we turn to \emph{invariant features}: moments of the signal which are invariant under shifts. We denote by $\hat x$ the discrete Fourier transform (DFT) of $x$, with $\hat{x} [k]= \sum_{n=0}^{L-1}x[n]e^{-2\pi i n k/L}$.
A shift by $r$ adds phase to the DFT: $\widehat{(R_{r}x)}[k] = \hat x[k]e^{-\frac{2\pi i r}{L} k}$. 
Using this fact, it is easy to verify that the mean, power spectrum, and bispectrum of $x$, defined by the formulae
\begin{align}
	\mu_x  & := \hat x[0] /L, \\  
	P_x[k] & := \hat x[k] \overline{\hat x[k]} = |\hat x[k]|^2,\\  
	B_x[k, \ell] & := \hat x[k] \overline{\hat x[\ell]} \hat x[\ell - k],
\end{align}
respectively, are invariant to shifts.

Simple expectation computations show the average invariant features of the measurements converge to the invariant features of the signal (up to bias terms) as $N \to \infty$, allowing to estimate them:
\begin{align}
	M_1 & := \frac{1}{N} \sum_{j = 1}^N \mu_{y_j} \to \mu_x, \label{eq:M1} \\
	M_2 & := \frac{1}{N} \sum_{j = 1}^N P_{y_j} \to P_x + \sigma^2 L \mathbf{1}, 
	 \label{eq:M2} \\
	M_3 & := \frac{1}{N} \sum_{j = 1}^N B_{y_j} \to B_x + \mu_x \cdot \sigma^2 L^2 A, \label{eq:M3}
\end{align}
where $\mathbf{1}$ is a vector of all-ones and $A \in \reals^{L\times L}$ is a zero matrix except for $A[0, 0] = 3$ and 1's on the remaining entries of the diagonal and the first row and column (if working with complex signals, subtract one from the first column of $A$).

The variance on $M_1$ scales like $O(\sigma^2/N)$. Because they square and cube the noise, variance on $M_2$ and $M_3$, respectively, scales like $O(\sigma^4/N)$ and $O(\sigma^6/N)$, {with cross-terms contributing additional $O(\sigma^2/N)$ variance, relevant only at high SNR}. Thus, provided $N/(\sigma^2+\sigma^6)$ is large enough (which is necessary for MRA~\cite{bandeira2017optimal}), the invariant features $\mu_x, P_x$ and $B_x$ can be estimated reliably. 
Various algorithms were proposed in~\cite{bendory2017bispectrum}  to recover $x$  from these moments. In the following section, we show how similar principles can be harnessed for the heterogeneous MRA model.


\section{Heterogeneity via mixed invariant features} \label{sec:mra_het}

In this work, we extend the invariant features approach to heterogeneous MRA. In this setup, $K$ unknown signals $x_1, \ldots, x_K \in \RL$ must be estimated (we assume they are distinct even up to shift), from the $N$ observations
\begin{align}
	y_j & = R_{r_j}x_{v_j} + \varepsilon_j, \quad j = 1, \dots, N, 
	\label{eq:mra_het}
\end{align}
where classes $v_j$ as well as shifts $r_j$ are unknown, and $\varepsilon_j$ is i.i.d.\ white Gaussian noise of variance $\sigma^2$ as before. We assume $v_j$'s are drawn i.i.d.\ 
from a (possibly unknown) mixing probability $w \in \Delta_K = \{ w\in\reals^K : w\geq 0 \textrm{ and } \sum_k w[k] = 1 \}$ (the simplex): $w[k]$ indicates the proportion of measurements which come from class $k$. 
Without the shifts, this model reduces to the well-studied Gaussian mixture model (GMM) with known, diagonal noise covariance, for which low-order moment methods have been studied~\cite{hsu2013learning,anandkumar2014tensor}.

Our goal in heterogeneous MRA is to esimate the signals $x_1, \ldots, x_K$ (up to shifts and ordering) and possibly to estimate $w$.  As before, we are not interested in $r_j$ or $v_j$ of individual measurements. 
We propose to do this based on $M_1, M_2$ and $M_3$~\eqref{eq:M1}--\eqref{eq:M3}, which are now \emph{mixed} invariant features. Expectation computations yield: 
\begin{align}
	M_1 & \to \sum_{k = 1}^K w[k] \mu_{x_k}, \\
	M_2 & \to \sum_{k = 1}^K w[k] P_{x_k} + \sigma^2 L \mathbf{1}, \\
	M_3 & \to \sum_{k = 1}^K w[k] \left(B_{x_k} + \mu_{x_k} \cdot \sigma^2 L^2 A\right).
\end{align}

Having computed $M_1, M_2$ and $M_3$ in $O(NL^2)$ flops (parallelized over $N$), we search for $K$ signals and possibly for a mixing density $w$ which best agree with the data in a least-squares sense.
The weights below proceed from a crude approximation of the variances of the individual terms, where it is assumed the power spectra of the unknown signals are roughly constant and close to the value $P$ (Appendix~\ref{apdx:weights})
(a common factor $\frac{N}{\sigma^2 L}$ was suppressed):
\begin{multline}
	\min_{\substack{\tilde x_1,\dots,\tilde x_K\in\RL \\ \tilde w\in\Delta_K}}    \left|\sum_{k=1}^{K} \tilde w[k] L\mu_{\tilde x_k} - LM_1\right|^2 \\ + 
	\frac{1}{\sigma^2 L + 2P} \left\|\sum_{k=1}^{K} \tilde w[k] P_{\tilde x_k} + \sigma^2L \mathbf{1} - M_2\right\|_\textrm{2}^2 \\ +  \frac{1}{\sigma^4 L^2 + 3P^2} \left\|\sum_{k=1}^{K} \tilde w[k] B_{\tilde x_k} + M_1 \cdot \sigma^2L^2 A - M_3\right\|_\textrm{F}^2.
\label{eq:opt}
\end{multline}
We simplified the third-order term somewhat by substituting $M_1$ for $\sum_k \tilde w[k] \mu_{\tilde x_k}$. The cost function is smooth in all variables, but it is non-convex. We use Manopt~\cite{manopt} to optimize it from random initializations. This toolbox allows to turn the simplex $\Delta_K$ into a Riemannian manifold~\cite{sun2015multinomial}, then to run a trust-region algorithm. 
The cost and its gradient can be computed in $O(KL^2)$ flops---independent of $N$. Because of non-convexity, the algorithm could converge to suboptimal points. In Section~\ref{sec:numerical}, we observe that this is rarely the case in practice for a wide range of parameters.

Importantly, our approach relies only on invariant features up to third order---a concise summary of the data as soon as $N \gg L$---and these can be estimated accurately as long as $N /(\sigma^2 + \sigma^6)$ is large enough. 
This in turn implies that the noise levels can be arbitrarily high, provided sufficiently many measurements are available.
 

\section{Numerical Experiments}\label{sec:numerical}

We conduct a few numerical experiments solving~\eqref{eq:opt}. In all our experiments, signals $x_1, \ldots, x_K$ are generated with i.i.d.\ standard Gaussian entries.
For two signals $x$ and $\tilde x$, we define a cyclic-shift invariant distance as
\begin{align}
	\dist(x, \tilde x) & = \min_{s \in \{0, \ldots, L-1\}} \| R_s x - \tilde x \|_2.
\end{align}
This is computed in $\calO(L\log L)$ flops with FFTs. An estimator $\mathbf{\tilde x} = (\tilde x_1, \ldots, \tilde x_K)$ for $\mathbf{x} = (x_1, \ldots, x_K)$ is defined up to ordering. Thus, we define the permutation invariant distance:
\begin{align}
	\dist(\mathbf{x}, \mathbf{\tilde x})^2 & = \min_{\pi \in S_K} \sum_{k = 1}^K \dist(x_k, \tilde x_{\pi(k)})^2.
	\label{eq:distpermutation}
\end{align}
Optimization over $S_K$ (permutations over $K$ elements) is  solved via the Hungarian algorithm in $\calO(K^3)$ operations.
The relative estimation errors we report below are given by:
\begin{align}
	\operatorname{relative\_error}(\mathbf{x}, \mathbf{\tilde x}) & = \frac{\dist(\mathbf{x}, \mathbf{\tilde x})}{\sqrt{\sum_{k = 1}^K \|x_k\|_2^2}}.
	\label{eq:metric}
\end{align}
If the mixing probabilities $w$ are estimated by $\tilde w$, given an optimal permutation $\pi$ in~\eqref{eq:distpermutation}, we report the estimation error as a total variation distance over the simplex:
\begin{align}
	\operatorname{TV\_dist}(w, \tilde w) & = \frac{1}{2} \sum_{k = 1}^K |w_k - \tilde w_{\pi(k)}|.
\end{align}
This value is between 0 and 1.

\begin{figure}[t]
	\begin{center}
	\includegraphics[width=1\columnwidth]{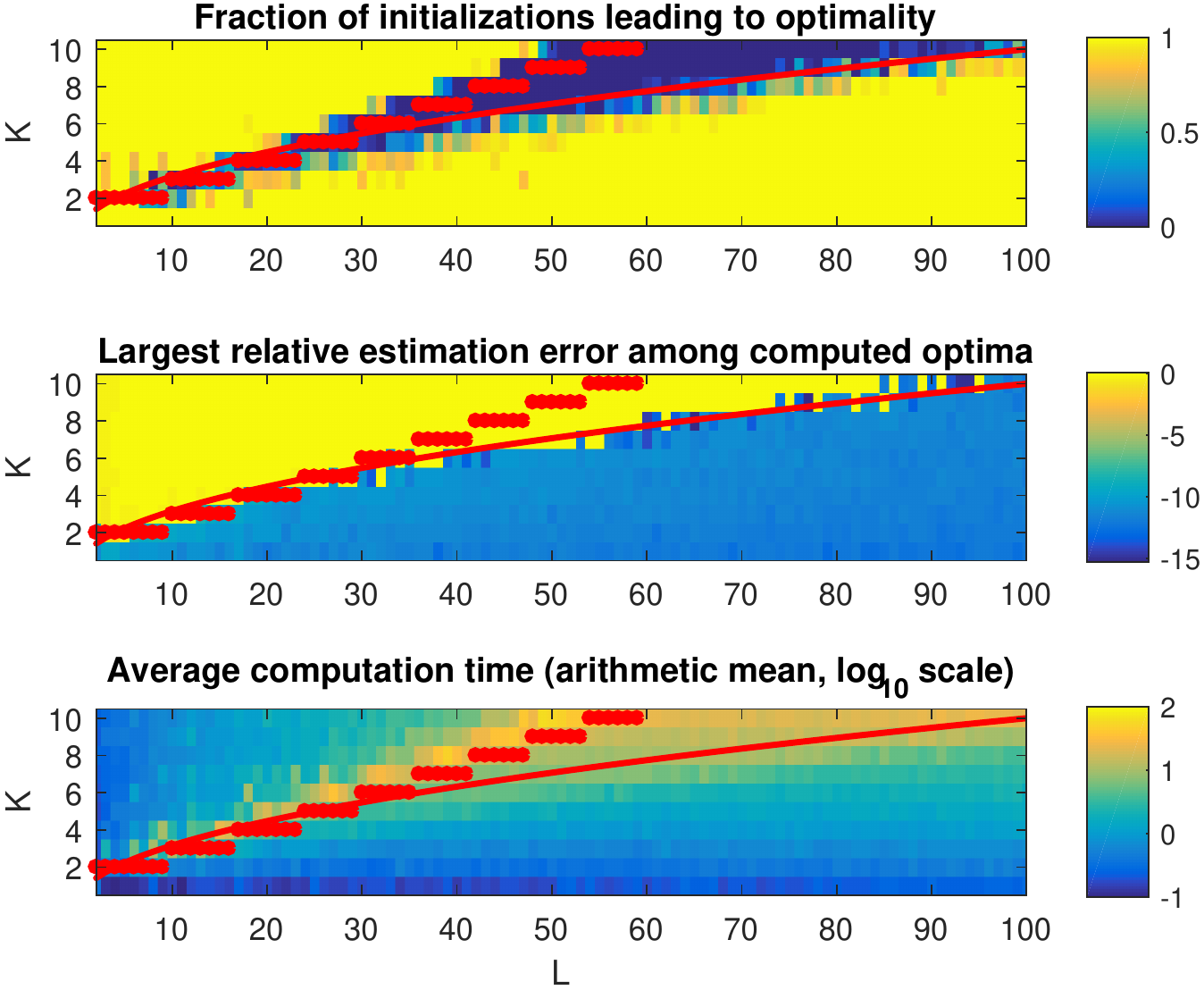}
	\caption{Experiment~1 ($N \to \infty$) suggests $K$ up to $\sqrt{L}$ (red curve) i.i.d.\ Gaussian signals can be demixed from perfect mixed invariant moments with~\eqref{eq:opt}. CPU time in seconds. Above red dots, recovery is hopeless because of an information theoretic argument. All but first plot in $\log_{10}$ scale. $L$ ranges from 2 to 100, $K$ from 1 to 10.}
	\label{fig:XP1}
	\end{center}
\end{figure}

\begin{figure}[t]
	\begin{center}
	\includegraphics[width=1\columnwidth]{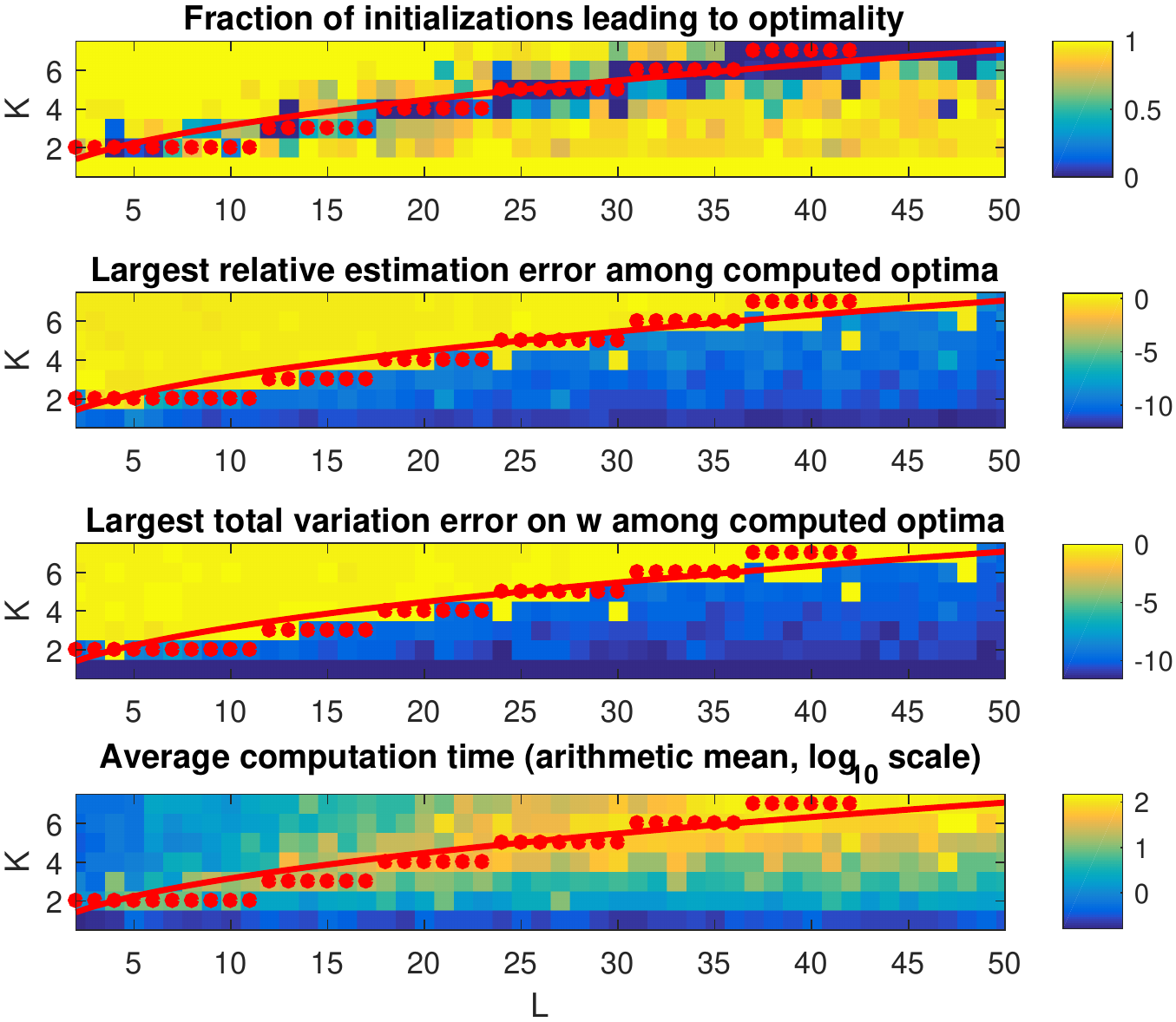}
	\caption{Experiment~2 ($N \to \infty$) is the same as Experiment~1, except mixing probabilities $w$ are now random and unknown.
		For the middle plots, yellow pixels inside the blue area correspond to setups where
		the smallest entry in $w$ is small, which is challenging as the corresponding signal is under-represented.  All but first plot in $\log_{10}$ scale. $L$ ranges from 2 to 50, $K$ from 1 to 7.}
	\label{fig:XP2}
	\end{center}
\end{figure}

\textbf{Red dots} on Figures~\ref{fig:XP1} and~\ref{fig:XP2} mark upper bounds on how large $K$ may be as a function of $L$ for demixing to be possible. We reason as follows: for generic real signals, $M_1$ provides 1 real number; $M_2$ provides $\sim \frac{1}{2}L$ distinct real numbers, and $M_3$ provides $\sim\frac{1}{6}L^2$ distinct real numbers (separating real and imaginary parts, and accounting for symmetries; a precise accounting is used for the figures.) A total of $KL$ real numbers must be estimated to recover the signals, and possibly an additional $K-1$ numbers are required to estimate mixing probabilities. Thus, even if all numbers provided by $M_1, M_2, M_3$ bear independent information, we still need $\sim\frac{1}{6}L^2$ to exceed $KL$ or $KL+(K-1)$. Solving for $K$ yields the displayed bound. In both cases, for large $L$, the bounds behave like $\lceil L/6 \rceil$. 

\textbf{Experiment~1} explores an infinite data regime, where the mixed invariant features $M_1, M_2, M_3$ are available exactly. The question is then: from these mixed features, can we recover the individual signals $x_1, \ldots, x_K$ by solving~\eqref{eq:opt}? If so, how large can we allow $K$ to grow as a function of $L$? Two separate issues are involved: (a) can we solve~\eqref{eq:opt} to global optimality despite non-convexity? And (b), do global optimizers coincide with the ground truth?
For values of $K$ and $L$ on a grid, we generate ground truth signals once with i.i.d. standard Gaussian entries (the power spectrum $P$ is the constant $L$ in expectation), and their exact mixed features are computed, with uniform mixing probability $w$ (known to the algorithm). Then, for each pair $(L, K)$, we generate 30 random initial guesses for the algorithm and optimize. We declare a global optimum is found if the cost value drops below $10^{-16}$. (In all experiments, the cost function is scaled by $\frac{\sigma^4L^2 + 3P^2}{2}$.) For each run where optimality is declared, we compute the relative estimation error according to~\eqref{eq:metric} and report the worst one for that $(L, K)$---the worst one, because in practice we would not be able to distinguish between global optima. 
Figure~\ref{fig:XP1} leads to the following empirical observation: (i) for $K$ up to approximately $\sqrt{L}$, the local optimization algorithm reliably finds a global optimizer, and it corresponds to the ground truth within numerical errors. There is also a regime where $K$ is so large that recovery is impossible, yet the optimization problem is easily solved.
In that regime, the solution to the problem is not unique: the problem is ill-posed.

\textbf{Experiment~2} is the same as Experiment~1, except the mixing probabilities $w$ are now random and unknown to the algorithm.
For values of $K$ and $L$ on a grid, we generate ground truth signals once, and a mixing probability $w$ as a vector whose entries are i.i.d.\ uniform in $[0, 1]$, normalized to be a probability density. Their exact mixed invariant moments are computed.
Figure~\ref{fig:XP2} suggests $w$ can also be recovered, although performance deteriorates.

\begin{figure}[t]
	\begin{center}
		\includegraphics[width=.8\columnwidth]{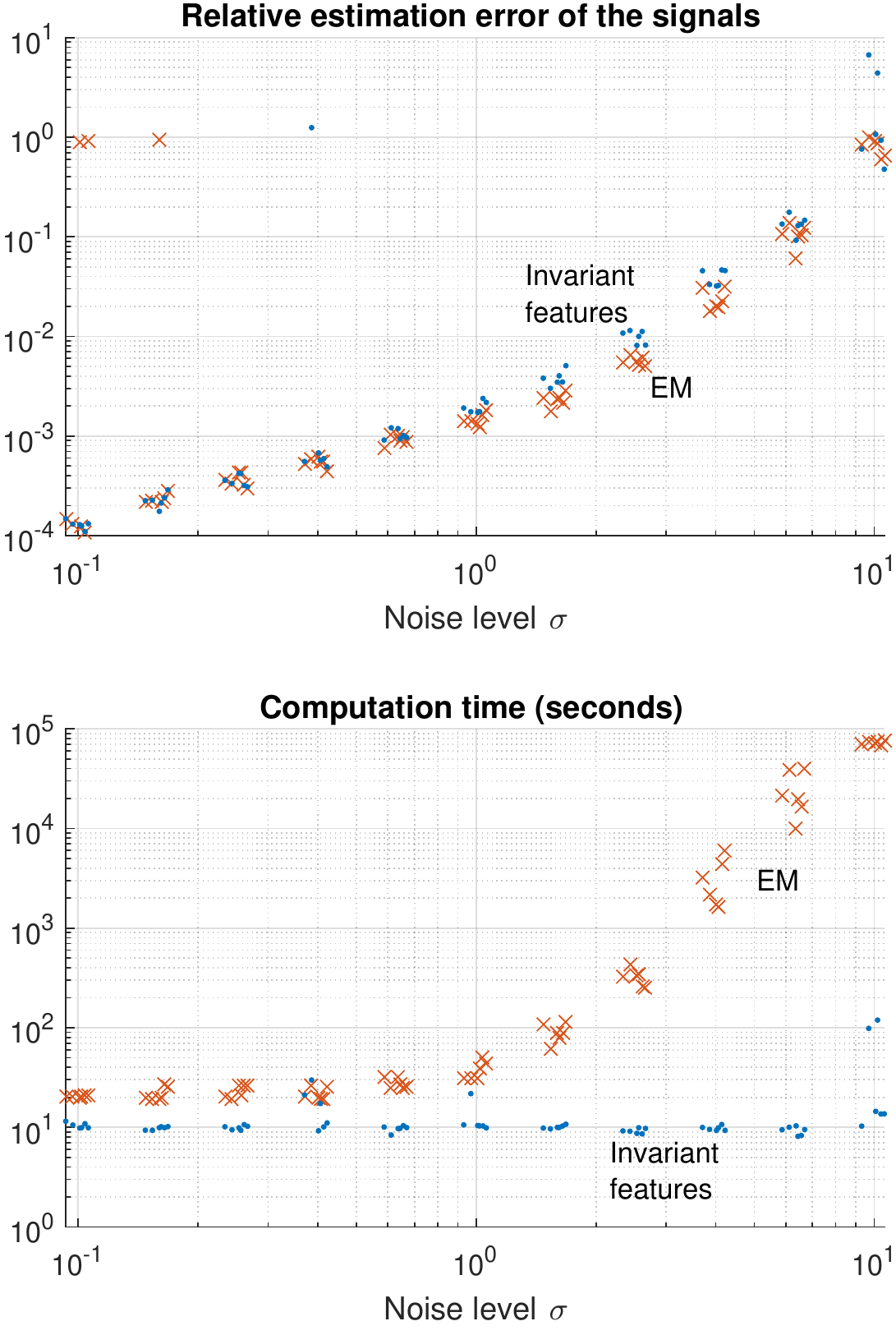}
		\caption{Experiment~3 ($L = 50, K = 2, N = 10^6$) shows both our algorithm (blue dots) and EM (red crosses) are resilient to high noise levels. EM is more accurate and our method is orders of magnitude faster. Jitter is added on the $\sigma$ axis for visualization. At $\sigma = 10$, EM attains its iteration limit of $10^4$.}
		\label{fig:XP3}
	\end{center}
\end{figure}

\textbf{Experiment~3} investigates resilience of the algorithm to high levels of noise---see Figure~\ref{fig:XP3}. We demix $K = 2$ signals of length $L = 50$ from $N = 10^6$ observations. Mixing probabilities $w$ are uniform. Our algorithm initializes $\tilde w$ uniform, but optimizes for it as well. For each value of noise level $\sigma$ on a logarithmic grid, signals are generated 6 times and two methods are run: (a) our algorithm based on~\eqref{eq:opt}, where we run the method with two random initial guesses and return the best result (based on attained cost value, which is available in practice); and (b) Expectation--Maximization (EM)---see below. We find that both methods are resilient to high noise, with EM producing more accurate estimators but our method being orders of magnitude faster for large noise.  

\textbf{EM} is a classic heuristic to address estimation problems with latent variables~\cite{dempster1977maximum}. In a nutshell (see~\cite{bendory2017bispectrum} for details in the homogeneous case), assuming a current estimate $\tilde x_1, \ldots, \tilde x_K$ is correct, EM computes $w_{j,r,k}$: the probability that measurement $y_j$ comes from signal $\tilde x_k$ with shift $r$. Concretely, under the Gaussian noise model, these probabilities are given by 
\begin{align}
	w_{j,r,k} \propto \exp\left( -\frac{\|R_r \tilde x_k - y_j\|_2^2}{2\sigma^2} \right),
\end{align}
and global scaling is fixed using $\sum_{r,k} w_{j,r,k} = 1$ for all $j$. Then, signal estimators are updated as follows: for each $k$,
\begin{align}
	\tilde x_k \leftarrow \frac{\sum_{j, r}w_{j,r,k} R_r^{-1} y_j}{\frac{\sigma^2}{\sigma_0^2} + \sum_{j, r}w_{j,r,k}},
	\label{eq:EMupdate}
\end{align}
where $\sigma_0 > 0$ comes from a prior on $x_k$'s coming from a distribution $\mathcal{N}(0, \sigma_0^2 I)$, where we pick $\sigma_0^2 = 1$ to match the true generating model, as a hint to EM. (Setting $\sigma_0^2 = 10^9$ to make the prior almost uninformative does not change the results much.)
The probabilities and the updated estimators can be computed in $O(NKL\log L)$ flops using FFTs, parallelizable over $N$ and $K$---we use Matlab's built-in parallelization of FFTs over $N$. We iterate until two subsequent iterates differ by less than $K \cdot 10^{-5}$ in metric~\eqref{eq:distpermutation}. (Results are robust against this choice.)

As Figure~\ref{fig:XP3} reveals, the number of EM iterations grows with the noise level (for $\sigma = 10^{-1}$, as little as 3 iterations suffice, while it saturates at our limit of 10\,000 for $\sigma = 10^1$). Attempts to reduce the strong effect of a large $N$ on the complexity of EM using batch iterations were not successful for this experiment. EM attains the most accurate estimators. One failure mode of EM is when the sum in the denominator of~\eqref{eq:EMupdate} (almost) vanishes for some signal: one of the estimators $\tilde x_k$ is given (almost) none of the observations, and this situation endures through iterations. Strangely, this occurs at high SNR, visible in Figure~\ref{fig:XP3}.

About parallelization: our method uses 30 cores to compute features $M_1, M_2, M_3$ in parallel over $N$---this is the bottleneck---then uses a single core to optimize from two random initial guesses sequentially (these could be done in parallel.) EM uses about 16 cores (number chosen by Matlab) to compute FFTs in parallel over $N$.
{Code: \scriptsize{\url{https://github.com/NicolasBoumal/HeterogeneousMRA}.}}

\section{Conclusions}\label{sec:conclusions}

We explored non-convex optimization for heterogeneous MRA, based on invariant features of order up to three. Key properties of our method are that (a) it never seeks to estimate the shifts or classes of the observations (which may be impossible to estimate at low SNR), and (b) once invariant features are computed (this takes linear time and is parallelizable in $N$), the complexity of the method no longer depends on the number of measurements $N$, which may need to be large for low SNR. Unfortunately, we are not able to theoretically explain the performance of the second stage at the moment.

In numerical experiments not shown here, with mixed moments known exactly (infinite data regime), we noticed that if we overestimate $K$ and allow the algorithm to optimize for $w$, then the algorithm still recovers the true signals, and assigns probability close to 0 to the extra estimated signals. If we underestimate $K$, then the algorithm tends to better estimate those signals which have larger probability $w[k]$. We intend to explore this further.

A key motivation for this work is the problem of heterogeneity in cryo-EM and in XFEL, both imaging techniques in structural biology. Heterogeneous MRA acts as a simplified model for those applications. Seminal work by Kam~\cite{kam1980reconstruction}, who showed how moment-based approaches for cryo-EM without heterogeneity can work, suggests our findings here may translate to handle heterogeneity in cryo-EM and XFEL.

\section*{Acknowledgments}

We gratefully acknowledge fruitful discussions with Nir Sharon, Will Leeb, Joe Kileel and Afonso Bandeira.

\bibliographystyle{ieeetr}
\bibliography{ref}


\clearpage

\appendix

\section{Weighing the cost function}\label{apdx:weights}

We pick the weights in the cost function of problem~\eqref{eq:opt} as follows. Along the way, we make a number of simplifying assumptions to keep formulas straightforward.
At the ground truth signal $\mathbf{x} = (x_1, \ldots, x_K) \in (\RL)^K$ and mixing probabilities $w \in \Delta_K$,
the error variables are as follows:
\begin{align*}
	& E_1 = M_1 - \sum_{k=1}^{K} w[k] \mu_{x_k}, \quad
	E_2 = M_2 - \sigma^2L \mathbf{1} - \sum_{k=1}^{K} w[k] P_{x_k}, \\
	& E_3 = M_3 - M_1 \cdot \sigma^2L^2 A - \sum_{k=1}^{K} w[k] B_{x_k}.
\end{align*}
By construction, they have zero mean.
If all error terms were (entry-wise) independent and Gaussian (neither is true\footnote{In particular, certain entries of the power spectrum and the bispectrum are repeated.}), then minimizing the sum of squared errors normalized by their individual variances would yield a maximum likelihood-type estimator.
This motivates us to normalize by (approximate) variances, as follows. If $y = x_{v} + \varepsilon$
(ignoring the shift $R_{r}$ since the features are invariant under it)
with $\varepsilon \sim \mathcal{N}(0, \sigma^2 I_L)$ and $v \sim w$ (independent), then by independence of the measurements,
\begin{align*}
	E_1 & = \frac{1}{N} \sum_{j = 1}^N \mu_{y_j} - \sum_{k=1}^{K} w[k] \mu_{x_k} \\
	    & = \frac{1}{N} \sum_{j = 1}^N \mu_{\varepsilon_{j}} + \left[\frac{1}{N} \sum_{j = 1}^N \mu_{x_{v_j}} - \sum_{k=1}^{K} w[k] \mu_{x_k}\right].
\end{align*}
The bracketed term is zero for homogeneous MRA. We neglect it for the heterogeneous case. Then, by independence,
\begin{align*}
	\Var\{E_1\} & \approx \frac{1}{N} \Var\{ \mu_{\varepsilon} \} = \frac{\sigma^2}{NL}.
\end{align*}
Proceeding similarly for $E_2$ we first get
\begin{align*}
	E_2[k] & = \frac{1}{N} \sum_{j = 1}^N |\hat y_j[k]|^2 - \sigma^2L - \sum_{k'=1}^{K} w[k'] |\hat x_{k'}[k]|^2 \\
	       & = \frac{1}{N} \sum_{j = 1}^N \left[|\hat x_{v_j}[k]|^2 + |\hat \varepsilon_j[k]|^2 + 2\Re\{ \overline{\hat x_{v_j}[k]} \hat \varepsilon_j[k] \} \right] \\
	       & \quad  - \sigma^2L - \sum_{k'=1}^{K} w[k'] |\hat x_{k'}[k]|^2 \\
	       & \approx \frac{1}{N} \sum_{j = 1}^N \left[|\hat \varepsilon_j[k]|^2 - \sigma^2L + 2\Re\{ \overline{\hat x_{v_j}[k]} \hat \varepsilon_j[k] \} \right],
\end{align*}
where again we neglected a term which vanishes exactly in the homogeneous case. The terms $|\hat \varepsilon_j[k]|^2$ and $\Re\{ \overline{\hat x_{v_j}[k]} \hat \varepsilon_j[k] \}$ are uncorrelated because $\hat \varepsilon_j[k]$ is distributed isotropically in the complex plane (in particular, the phase is uniform.) Thus, we can separate the variance computation in two parts:
\begin{align*}
	\Var\{ E_2[k] \} & \approx \frac{1}{N} \Var\{ |\hat\varepsilon[k]|^2 \} + \frac{1}{N} \Var\{ 2 \Re\{ \overline{\hat x_{v}[k]} \hat \varepsilon[k] \} \}.
\end{align*}
We can easily understand the distribution of $\hat \varepsilon$. Indeed, let $F$ be the DFT matrix so that $\hat \varepsilon = F \varepsilon$ for $\varepsilon \sim \mathcal{N}(0, \sigma^2I_L)$. Noting that $FF^* = LI_L$, we get: $\EE\{ \hat\varepsilon \hat \varepsilon^* \} = F\EE\{\varepsilon \varepsilon^*\}F^* = \sigma^2 L I_L$. Thus, $\hat \varepsilon \sim \mathbb{C}\mathcal{N}(0, L\sigma^2 I_L)$.
Let $\hat \varepsilon[k] = z_1 + iz_2$, with $z_1, z_2 \sim \mathcal{N}(0, \frac{\sigma^2 L}{2})$. Then,
\begin{align*}
	\Var\{ |\hat\varepsilon[k]|^2 \} & =  \Var\{ z_1^2 + z_2^2 \} = \sigma^4 L^2.
\end{align*}
On the other hand, using $\EE\{ \hat \varepsilon[k]^2 \} = 0$ due to uniform phase again and independence of $\varepsilon$ and $v$,
\begin{multline*}
	\Var\{ 2 \Re\{ \overline{\hat x_{v}[k]} \hat \varepsilon[k] \} \} = \EE\{ ( \overline{\hat x_{v}[k]} \hat \varepsilon[k] + \hat x_{v}[k] \overline{\hat \varepsilon[k]} )^2 \} \\
			= 2\EE\{ |\hat x_{v}[k]|^2\} \EE\{ |\hat \varepsilon[k]|^2  \}
			 = 2\sigma^2 L \sum_{k'=1}^K w[k'] |\hat x_{k'}[k]|^2.
\end{multline*}
Overall,
\begin{align*}
	\Var\{ E_2[k] \} & \approx \frac{\sigma^2 L}{N}\left( \sigma^2 L + 2 \sum_{k'=1}^{K} w[k'] P_{x_{k'}}[k] \right).
\end{align*}
The sum is nothing but the mixed power spectrum, which we can estimate from the data: this could be used as weight directly. To simplify even further, assuming the power spectra of the signals are not too far from the constant $P$ (in the experiments, $x$ has i.i.d.\ standard entries, so the power spectrum is close to $L$), we can approximate the variance as a constant:
\begin{align*}
	\Var\{ E_2[k] \} & \approx \frac{\sigma^2 L}{N}\left( \sigma^2 L + 2P \right).
\end{align*}
We now turn to the bispectrum: each measurement $y_j$ contributes eight terms to $E_3[k, \ell]$ through $\hat y_j[k] \overline{\hat y_j[\ell]} \hat y_j[\ell-k]$ and $\hat y_j = \hat x_{v_j} + \hat  \varepsilon_{j}[k]$ (again, ignoring the shift $R_{r_j}$ since the bispectrum is invariant under it.) In approximating the variance, we aim to identify the leading terms in $\sigma$ for low and for high SNR. As for $E_1$ and $E_2$, a term independent of $\sigma$ which vanishes exactly in the homogeneous case is neglected here. Thus, it remains to identify the terms which scale as $\sigma^2$ and $\sigma^6$. These come from:
\begin{align*}
	\Var\{\hat \varepsilon[k] \overline{\hat \varepsilon[\ell]} \hat \varepsilon[\ell-k]\}
\end{align*}
for $\sigma^6$ and from
\begin{align*}
	& \Var\{\hat x_{v}[k] \overline{\hat x_{v}[\ell]} \hat \varepsilon[\ell-k]\}, \\
	& \Var\{\hat x_{v}[k] \overline{\hat \varepsilon[\ell]} \hat x_{v}[\ell-k]\}, \\ 
	& \Var\{\hat \varepsilon[k] \overline{\hat x_{v}[\ell]} \hat x_{v}[\ell-k]\}
\end{align*}
for $\sigma^2$. Aiming for a crude approximation, for the $\sigma^6$ term we consider only the case where $\varepsilon[k], \varepsilon[\ell], \varepsilon[\ell-k]$ are independent, in which case
\begin{align*}
	\Var\{\hat \varepsilon[k] \overline{\hat \varepsilon[\ell]} \hat \varepsilon[\ell-k]\} & = \EE\{ |\hat \varepsilon[k]|^2 \}^3 = (\sigma^2 L)^3.
\end{align*}
(For values of $k, \ell$ where independence does not hold, an extra constant would appear.) For the $\sigma^2$ terms, the first one expands as
\begin{align*}
	\Var\{\hat x_{v}[k] \overline{\hat x_{v}[\ell]} \hat \varepsilon[\ell-k]\} & = \EE\{|\hat x_{v}[k]|^2 |\hat x_{v}[\ell]|^2\} \EE\{|\hat \varepsilon[\ell-k]|^2\}.
\end{align*}
The first expectation (over $v$) can be estimated from the data: it is a mixture of fourth order moments. Unfortunately, estimating it accurately would require $\calO(\sigma ^8)$ observations. Alternatively, for the homogeneous case, it could be approximated using the estimated power spectrum. Simpler still, as we do here, assuming the power spectra of the signals to be estimated are close to $P$, we approximate
\begin{align*}
	\Var\{\hat x_{v}[k] \overline{\hat x_{v}[\ell]} \hat \varepsilon[\ell-k]\} & \approx \sigma^2 L P^2.
\end{align*}
There are three such terms, so that overall we get the approximation:
\begin{align*}
	\Var\{ E_3[k, \ell] \} & \approx \frac{\sigma^2 L}{N}\left( \sigma^4 L^2 + 3P^2 \right).
\end{align*}

\end{document}